\documentclass[]{JHEP3}

\JHEPspecialurl{http://jhep.sissa.it/JOURNAL/JHEP3.tar.gz}

\usepackage{graphicx}
\usepackage{amsmath}
\usepackage{epsfig,multicol,bbm}
\usepackage{slashed}

\newcommand{\be}{\begin{equation}}
\newcommand{\ee}{\end{equation}}
\newcommand{\bea}{\begin{eqnarray}}
\newcommand{\eea}{\end{eqnarray}}

\newcommand{\half}{\frac{1}{2}}
\newcommand{\mc}{\mathcal}

\newcommand{\beqa}{\begin{eqnarray}}
\newcommand{\eeqa}{\end{eqnarray}}

\newcommand{\nn}{\nonumber}

\def\sl{\slashed}

\def\beq{\begin{equation}}
\def\eeq{\end{equation}}

\newcommand\fverb{\setbox\fverbbox=\hbox\bgroup\verb}
\newcommand\fverbdo{\egroup\medskip\noindent%
			\fbox{\unhbox\fverbbox}\ }
\newcommand\fverbit{\egroup\item[\fbox{\unhbox\fverbbox}]}
\newbox\fverbbox

\title{Supersymmetric Radiative Flavour}

\author{Joseph P. Conlon, Francisco G. Pedro\\
	Rudolf Peierls Centre for Theoretical Physics, University of Oxford, 1 Keble Road, Oxford, OX1 3NP, UK\\
	E-mail: \email{j.conlon1@physics.ox.ac.uk}, \email{f.pedro1@physics.ox.ac.uk}}

\preprint{OUTP-11-47-P}	

\abstract{
We examine possibilities for the radiative generation of the Yukawa couplings and flavour structure in 
supersymmetric models in the supersymmetric phase. 
Not withstanding the non-renormalisation of the Wilsonian superpotential,
this can occur through the 2-loop vertex renormalisation of the physical 1PI couplings.
We describe this effect and construct models in which this occurs.
For models attempting to reproduce the full flavour structure of the Standard Model, we analyse the tension between
such models and constraints from low-energy flavour observables. We note that the tension is weakest for the case of generating Dirac neutrino masses.
}


\begin{document}

\section{Introduction}

Understanding the flavour structure of the Standard Model is one of the outstanding problems of
theoretical physics. The pattern of masses and mixings associated to the quark and lepton sectors
shows a clear structure but has no definitive explanation.

One aspect of this structure is the presence of a heavy third generation together with two generations
that are much lighter, being approximately massless compared to the heavy generation. This motivates the idea
that the masses of the light generations are radiatively generated from the heavy third generation.

Furthermore, in the context of top-down models such as those in string theory,
it is often the case that superpotential Yukawa couplings show either
a rank-one or a rank-two flavour structure that is exact in perturbation theory.
This restricted rank structure has appeared for heterotic models \cite{0601204, 09042186}, 
IIA and IIB intersecting brane models \cite{0302105, 0404229}, IIB models of branes
at singularities \cite{aiqu,Krippendorf:2010hj} and F-theory models \cite{Cecotti:2009zf, 09102413}.
It is important to understand the physics that can generate rank three Yukawa couplings. Suggestions include
nonperturbative instanton corrections to the superpotential \cite{hepth0612110, 09105496} and radiative effects
associated to supersymmetry breaking \cite{11021973}.

In this paper we consider a mechanism that has received little to no attention. This is the possibility
that Yukawa couplings can be generated radiatively but supersymmetrically. Several supersymmetric
models have been considered in which radiative flavour generation
occurs as a result of SUSY breaking, for example \cite{11021973,Ibanez1982, 9601262, 08041996, 09064657}.
The novelty here is that we examine the possibility
that radiative flavour generation occurs even in the supersymmetric phase,
in the absence of supersymmetry breaking.

This possibility has been less considered as naively it is impossible - 
the superpotential is not renormalised; only the K\"ahler potential is renormalised,
and this cannot affect the rank of the superpotential. The loophole in this argument
 requires a subtle but not widely appreciated aspect of supersymmetric field theory \cite{West1, West2, JackJonesWest, DunbarJackJones},
   and in particular
 a careful distinction between the Wilsonian and 1PI actions.
Although the Wilsonian superpotential is not renormalised, the physical couplings are determined by the 1PI action.
The non-renormalisation theorems do not prevent radiative generation of new holomorphic couplings in the 1PI action.
For a string computation of this effect, see \cite{CGP}.

 The structure of the paper is as follows. In section 2 we review the physics of this effect.
 We explain how Yukawa couplings can be radiatively generated in a manner
 consistent with supersymmetry and describe the explicit evaluation of the relevant Feynman diagrams. In section 3 we consider
 extensions of the MSSM and potential applications of the effect 
 to generating the flavour structure for quarks, leptons and neutrinos. In section 4 we consider modifications that will occur once
 supersymmetry is broken, and in section 5 we consider the tension with low energy flavour observables.

\section{Wilsonian and 1PI Superpotentials}

In supersymmetric field theory
non-renormalisation theorems play an important role for both model building and formal aspects.
Certain terms - desirable or undesirable - consistent with the symmetries may be absent from the tree level superpotential.
If this is the case, the nonrenormalisation theorem says that their (perturbative) radiative generation is forbidden.

In this section we briefly review the precise statement of the non-renormalisation theorem and
explain the relevant loophole. It is possible to generate holomorphic couplings through radiative effects
 when interacting massless particles are present in the theory. This arises from the
distinction between the Wilsonian action, which is protected by holomorphy, and the 1PI action, which is not.

The proof of the non-renormalisation theorem can be found in \cite{Grisaru:1979wc}, where supersymmetric perturbation theory was used.  Seiberg, avoiding the technicalities of super Feynman rules, presented in  \cite{Seiberg:1993vc}  and  \cite{Seiberg:1994bp} an alternative derivation. Since we are mostly interested in perturbative effects we will follow \cite{Grisaru:1979wc} and state the non renormalisation theorem as follows:
All contributions to the Wilsonian effective action from loop diagrams are of the form
\be
\label{eq21}
A_G=\int d^ 4\theta d^4 x_1...d^4 x_n \, G(x_1,...,x_n, \theta,\bar{\theta}),
\ee
and involve an integral over all of superspace. In particular, no contributions arise of the form
\be
\label{eq22}
A_W=\int d^ 2\theta d^4 x_1...d^4 x_n \, W(x_1,...,x_n, \theta)+ c.c. ,
\ee
where the integral is only over the chiral subspace of superspace.

The key to understanding the limitations of the non-renormalisation theorem is to note that it
applies to the Wilsonian action. As pointed out in e.g. \cite{Shifman:1986} there are two objects that go under the designation of effective action: the Wilsonian effective action and the one particle irreducible effective action.
The Wilsonian action is defined at a particular scale $\mu$ by integrating out degrees of freedom with $E > \mu$ and retaining degrees of freedom with
$E < \mu$. As the energy scale $\mu$ is decreased the terms in the bare Lagrangian are modified in such a way as to keep long distance physics
unaltered. Terms in the Wilsonian action defined at a scale $\mu$ do not directly give the physical couplings, as 
for processes at a scale $E < \mu$ there still remain light degrees of freedom to integrate over. Physical couplings are instead given by the 1PI action, which includes the effect
of integrating over light fields. If all fields are massive, then it is possible to decouple all degrees of freedom by taking $\mu$ to be sufficiently small, and in this case the Wilsonian action is equivalent to the 1PI action.

However if there are massless degrees of freedom these remain light for all values of $\mu$ and to obtain the physical
 couplings it is necessary also to integrate
explicitly over the massless degrees of freedom. The 1PI action takes into account loops involving massless states, whereas the Wilsonian action does not. Such loops can give radiative vertex corrections.
The non-renormalisation theorem is the statement that the 
Wilsonian superpotential - not the 1PI superpotential - is not renormalised. However, as argued above,
in the presence of massless particles, this does not imply the non-renormalisation of the physical 
vertex couplings.\footnote{Another place where the distinction between Wilsonian and 1PI actions is often made is that of gauge couplings and the NSVZ beta function \cite{Shifman:1986, KL}. However in this case it is better to think instead of two Wilsonian actions, with one involving the conventional supersymmetric form of the action (with a gauge kinetic function) and one involving canonically normalised gauge fields. The NSVZ beta function arises not as a true infrared
effect but rather as the anomaly in the functional integral measure when
rescaling the gauge fields to canonical normalisation \cite{ArkaniHamedMurayama}. In contrast the physics we are interested in involves
genuinely infrared effects, requiring the presence of massless particles.}

It is worth emphasising that this is distinct from conventional wavefunction renormalisation, which also alters physical couplings.
Wavefunction renormalisation is easily captured in a Wilsonian action as a renormalisation of the K\"ahler potential. The effect described
here corresponds to a renormalisation of the interaction vertex rather than the external propagator legs.

It is in fact true that physical holomorphic couplings are indeed renormalised.
This effect is not unknown and was pointed out in \cite{West1, West2, JackJonesWest},
although it does not appear to have wide circulation. For example, the massless Wess-Zumino model has a 2-loop renormalisation of the $\phi^3$ vertex \cite{JackJonesWest}. In the languages of equations (\ref{eq21}) and (\ref{eq22}) the effect arises through the presence of a term
\be
\int d^4 x \, d^4 \theta \, \frac{1}{\square} D^2 g(\Phi) + c.c \;,
\ee
where $g(\Phi)$ is a holomorphic function of the chiral superfield $\Phi$. Replacing $\int d^4 \theta$ by $\int d^2 \theta \bar{D}^2$ and using $\bar{D}^2 D^2 \Phi = \square \Phi$, this gives an effective contribution
\be
\int d^4 x \, d^2 \theta  \, g(\Phi) \;,
\ee
acting as an effective superpotential operator. This requires the presence of massless particles, as otherwise the loop inverse propagator
$ \frac{1}{\square} \to \frac{1}{\square + m^2}$ and the contribution decouples at zero momentum.
In section 3 we describe the explicit evaluation of the effect using Feynman diagrams.

Renormalisation of holomorphic Yukawa couplings via superpotential couplings requires the diagram to be consistent with the chiral nature of superpotential interactions. This requires a 2-loop diagram as 1-loop diagrams are not consistent with chirality. This is easy to see
 diagrammatically as in figure 1.
 \begin{figure}[ht]
 \label{OneGraph}
\begin{center}
\includegraphics[width=10cm]{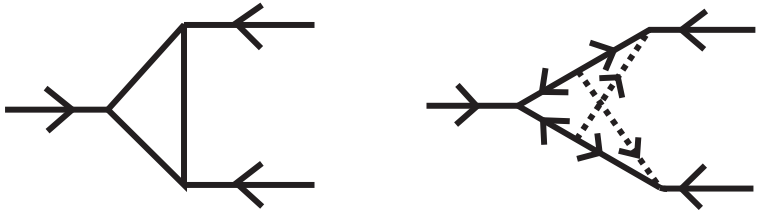}
\caption{For massless loop particles
it is impossible to draw a 1-loop diagram consistent with the chiral nature of interaction, which
requires vertices to be either holomorphic (all arrows ingoing) or antiholomorphic (all arrows outgoing).
This can however be done at two loops, as exemplified by the right hand diagram.}
\end{center}
\end{figure}

In the presence of gauge-charged particles there is a 1-loop diagram which
leads to Yukawa vertex renormalisation. However this involves the gauge fields and can
only renormalise existing couplings - it is unable to generate new couplings and so is less interesting for our purposes.

\section*{Evaluation of Graphs}

\begin{figure}[ht]
\begin{center}
\includegraphics[width=14cm]{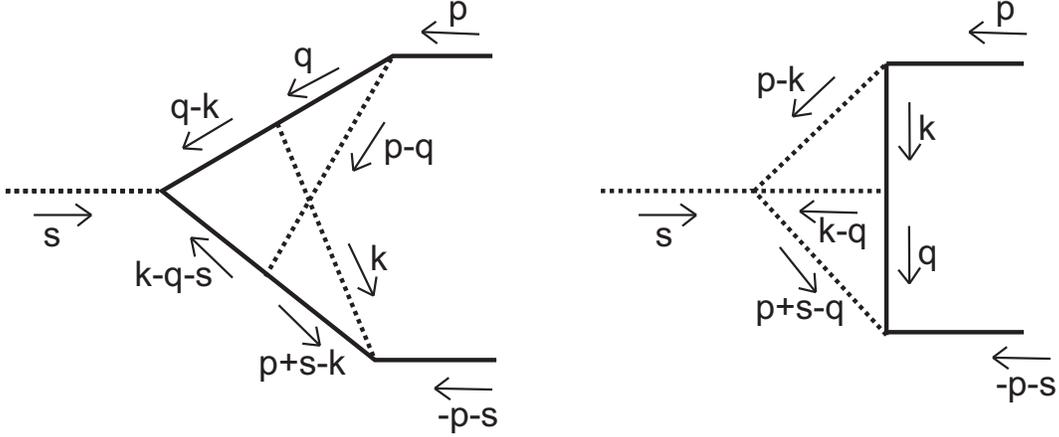}
\caption{The two graphs that contribute to the Yukawa vertex renormalisation. Dotted lines represent scalars and
solid lines fermions.}
\end{center}
\label{TwoGraphs}
\end{figure}

In this section we review the calculation of \cite{JackJonesWest} for the massless Wess-Zumino model
and fill in some details. The K\"ahler potential is $K = \Phi^{\dagger} \Phi$ and the
superpotential $W = \frac{\lambda}{6} \Phi^3$. This has a Yukawa
vertex $\lambda \phi \psi \psi$ and a scalar 4-point interaction $\frac{\lambda \lambda^{*}}{4} (\phi \phi^{*})^2$.
It is then possible to compute the 2-loop renormalised Yukawa directly.
There are two basic graphs to be considered, which are shown in figure \ref{TwoGraphs}.

With momenta labelled as shown, the first graph has an amplitude
\be
\label{amp1}
\mc{A}_1 = (\lambda^{*} \lambda)^2 \lambda
\int \frac{d^4 q}{(2 \pi)^4} \int \frac{d^4 k}{(2 \pi)^4}
\frac{\bar{u}(p) \slashed{q} (\sl{q} - \sl{k}) (\sl{s} + \sl{q} - \sl{k})(\sl{p} + \sl{s} - \sl{k})
v(-p-s)}{q^2 (q-k)^2 (s+q-k)^2 (p+s-k)^2 k^2 (p-q)^2}.
\ee
The second graph has an amplitude of
\be
\label{amp2}
\mc{A}_2 =
(\lambda^{*} \lambda)^2 \lambda
\int \frac{d^4 q}{(2 \pi)^4} \int \frac{d^4 k}{(2 \pi)^4}
\frac{\bar{u}(p) \sl{k} \sl{q} v(-p-s)}{k^2 q^2 (k-q)^2 (p-k)^2 (s+p-q)^2}.
\ee
To consider amplitudes with zero external momentum, we put $s=0$. Then equation (\ref{amp1}) becomes,
extracting the overall factor of $\bar{u}(p) v(-p)$,
\be
\frac{(\lambda^{*} \lambda)^2 \lambda}{(2 \pi)^8}
\int \frac{d^4 q \, d^4 k}{k^2 q^2}
\frac{\sl{q}(\sl{q}-\sl{k})(\sl{q} - \sl{k})(\sl{p} - \sl{k})}{(q-k)^4 (p-k)^2 (p-q)^2}
= \frac{(\lambda^{*} \lambda)^2 \lambda}{(2 \pi)^8}
\int \frac{d^4 q \, d^4 k}{k^2 q^2} \frac{\sl{q} (\sl{p} - \sl{k})}{(q-k)^2 (p-k)^2 (p-q)^2}.
\label{amp1res}
\ee
Eq. (\ref{amp2}) likewise becomes
\be
\label{amp2res}
\mc{A}_2 = \frac{(\lambda \lambda^{*})^2 \lambda}{(2 \pi)^8} \int d^4 k \, d^4 q \, \frac{ \sl{k} \sl{q}}{ k^2 q^2 (k-q)^2 (p-k)^2 (p-q)^2}.
\ee
Summing amplitudes (\ref{amp1res}) and (\ref{amp2res}) we get
\be
\mc{A} = \mc{A}_1 + \mc{A}_2 = \frac{(\lambda \lambda^{*})^2 \lambda}{(2 \pi)^8} \int d^4 k \, d^4 q  \, \frac{ \sl{q} \sl{p} }{ k^2 q^2 (k-q)^2 (p-k)^2 (p-q)^2}.
\ee
To simplify this expression we write $q^{'} = p - q$, $k^{'} = p - k$. As the loop variables are dummy variables,
we then obtain
\be
\mc{A} = \half p^2 \frac{\lambda (\lambda \lambda^{*})^2}{(2 \pi)^8} \int d^4 k \, d^4 q \, \frac{1}{(p-k)^2 (p-q)^2 (k-q)^2 k^2 q^2}.
\ee
We want to evaluate the two-loop integral
\be
\label{Zint}
Z = \int \frac{ d^n k \, d^n q}{q^2 k^2 (k+p)^2 (q+p)^2 (k-q)^2},
\ee
for the case $n=4$.
The derivation of this can be found in \cite{Tkachov:1981wb} or \cite{Jones:1982zf}. For completeness we will sketch the
necessary steps here. The integral $Z$ can be re-expressed in terms of 1-loop integrals $I(p^2)$ and $J(p^2)$,
where
\bea
I(p^2) & = & \int \frac{d^n k}{k^2 (k+p)^2}, \\
J(p^2) & = & \int \frac{d^n k}{(k^2)^{3-n/2} (k+p)^2}.
\eea
In $n$ dimensions, $Z$, $I$ and $J$ are related by (putting $p^2 = 1$)
\be
\label{ZIJ}
(n-4)Z = (6n-20) I J - (2n - 6)I^2.
\ee
$I$ and $J$ can be evaluated straightforwardly (e.g. as in \cite{PeskinSchroeder}) to be
\bea
I & = & \pi^2 \Gamma ( 2- \frac{n}{2}) B \left(\frac{n}{2} - 1, \frac{n}{2} - 1 \right) \left(p^2 \right)^{\frac{d}{2} - 2}, \\
J & = & \pi^2 \frac{\Gamma(4-n)}{\Gamma(3 - \frac{n}{2})} B\left(n-3, \frac{n}{2} - 1 \right) \left(p^2 \right)^{(d-4)}.
\eea
Using (\ref{ZIJ}) and the expansions of the gamma functions it follows that for $n = 4 + 2 \epsilon$
\be
\label{Zexp}
Z \vert_{\epsilon \to 0} \to 6 \pi^4 \zeta(3).
\ee

The derivation of the relation (\ref{ZIJ}) goes as follows. Consider the expression
$$
\int d^n p \, d^n q \, \frac{\partial}{\partial p_{\mu}}   \left( \frac{(p-q)_{\mu}}{p^2 (p-k)^2 q^2 (q-k)^2 (p-q)^2} \right).
$$
This vanishes as we integrate over a total derivative. However, by explicitly performing the derivative
and rewriting integration variables as $p \to p - k$, $q \to q-k$, we obtain
\be
0 = \int \, d^n p \, d^n q \frac{n-2}{p^2 (p-k)^2 q^2 (q-k)^2 (p-q)^2} - \frac{4 p \cdot (p-q)}{(p^2)^2 (p-k)^2 q^2 (q-k)^2 (p-q)^2}.
\ee
Using $2 p \cdot (p-q) = p^2 + (p-q)^2 - q^2$, this gives
\bea
\int d^n p \, d^n q \, \frac{4-n}{p^2 (p-k)^2 q^2 (q-k)^2 (p-q)^2}
& = & \int d^n p \, d^n q \, \frac{-2}{(p^2)^2 (p-k)^2 q^2 (q-k)^2} \nn \\
& & + \int d^n p \, d^n q \, \frac{2}{p^2 (p-k)^4 q^2 (p-q)^2}.
\eea
This factorises the original integral (\ref{Zint}) into two separate one-loop integrals, which can be evaluated by first
performing the $\int d^4 q$ integral and subsequently the $\int d^4 p$ integral. Doing so gives the relation
(\ref{ZIJ}).

Combining equations (\ref{Zint}) and (\ref{Zexp}) renormalises the Yukawa coupling $\lambda$ as
\be
\label{yukrenorm}
\lambda \to \lambda \left( 1 + 3 \zeta(3) \frac{(\lambda \lambda^{*})^2}{(4 \pi)^4} \right).
\ee

The above has described a component field calculation of the renormalised Yukawa coupling.
The calculation can also be carried out
using supergraphs in a manifestly supersymmetric fashion \cite{Buchbinder1, Buchbinder2}.
As the coupling (\ref{yukrenorm}) comes from a supersymmetric theory,
supersymmetry implies there is also a 2-loop renormalisation of the $\left( \phi \phi^{*} \right)^2$ vertex from
\be
\label{scalar2loop}
V_F = \left\vert \frac{\partial W}{\partial \Phi} \right\vert^{2}_{\phi} \to \lambda \lambda^{*}
\left(1 + 6 \zeta(3) \frac{(\lambda \lambda^{*})^2}{(4 \pi)^4} \right) (\phi \phi^{*})^2.
\ee
However in this case many more graphs can contribute, some of which are shown in figure 3.
\begin{figure}[ht]
\begin{center}
\includegraphics[width=14cm]{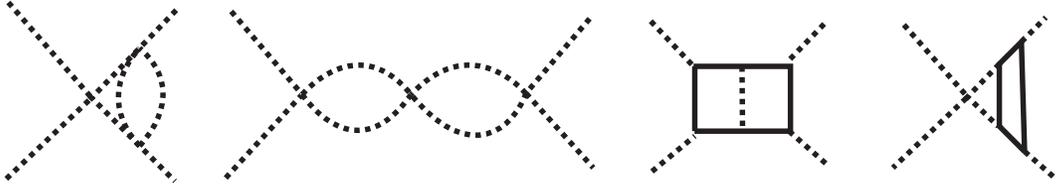}
\caption{Some of the graphs that can contribute to the 2-loop renormalisation of the scalar 4-point interaction.}
\end{center}
\label{ScalarGraphs}
\end{figure}
We expect the component field evaluation of (\ref{scalar2loop}) to be substantially more difficult and we do not attempt it.

For the more complicated models considered below the same two basic Yukawa renormalisation
diagrams of figure 2 apply. The first follows directly from the
superpotential and only uses the fermion-fermion-scalar trilinear coupling, 
while the second also utilises the 4-point scalar vertices from the F-term potential.

At higher loop further contributions to the effective chiral superpotential can be generated. Let us describe one
calculable example. The presence of a 2-loop correction to the superpotential, generated radiatively from the tree level
superpotential, suggests that that there should be a further 4-loop contribution, generated radiatively from the
2-loop term. Assuming the existence of the 2-loop correction to the scalar 4-point function 
in eq. (\ref{scalar2loop}), this correction is calculable.
Consider the 4-loop graphs shown in figure 4.
\begin{figure}[ht]
\begin{center}
\includegraphics[width=12cm, height=8cm]{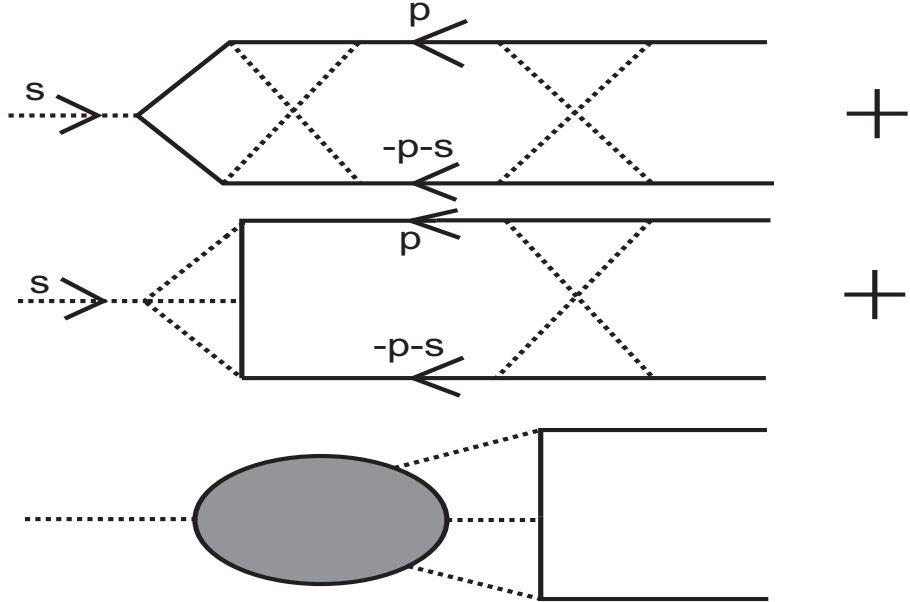}
\caption{A 4-loop correction to the chiral superpotential coming from a repeated iteration of the 2-loop
correction.}
\end{center}
\label{4loop}
\end{figure}
The first two graphs have on their left hand side a structure that is precisely the same as that found in the 2-loop case.
When $s=0$, the explicit evaluation of the 2-loop graphs leading to (\ref{yukrenorm}) in fact shows that the loop integral gives a finite value that
is independent of $p$. The first two graphs of figure 4 therefore give the same as the first graph of figure \ref{TwoGraphs}, multiplied
by an overall factor of $\left(3 \zeta(3) \frac{(\lambda \lambda^{*})^2}{(4 \pi)^4}\right)$. This will combine with the analogue of the
second graph of figure \ref{TwoGraphs}.
Assuming that the scalar 4-point function does receive the 2-loop correction required by supersymmetry, this is provided by
 the third graph of figure 4. The 4-loop graph then entirely factorises into two 2-loop integrals, giving an overall magnitude of
$\left(3 \zeta(3) \frac{(\lambda \lambda^{*})^2}{(4 \pi)^4}\right)^2$. Effectively, what has happened is that we have simply redone the
original computation of the 2-loop correction to the superpotential, except now treating 
the previously radiative term as the `tree-level' term.

We have given particular prominence to the above 4-loop term as it will allow a definite generation of a radiative first generation
Yukawa in some of the models discussed below. However we do not claim that this exhausts the structure of loop corrections
to holomorphic couplings. For example,
it is easy to write down diagrams that give potential 3-loop vertex corrections to the Yukawa couplings. However 
attempts to evaluate such diagrams explicitly is beyond the scope of this paper.

\section{Supersymmetric Models of Radiative Flavour}

We now investigate whether the effect described above can be used as a model for radiative
flavour generation in the Supersymmetric Standard Model. As described, the effect holds for unbroken supersymmetry and will be
modified once supersymmetry is broken.
In sections \ref{subsecquarks} and \ref{subsecleptons} we first
analyse radiative flavour generation in the unbroken MSSM
before considering in section 4
the necessary modifications once supersymmetry is broken.

For the Wess-Zumino model considered in section 2, the effect appeared as a renormalisation of an existing Yukawa
coupling. We first show briefly that it is possible to generate new Yukawa couplings that had vanished at tree level.
Consider a theory with 3 uncharged chiral superfields: $H$, $\Phi_2$ and $\Phi_3$.
For the superpotential we take\footnote{The superpotential has an R-symmetry under which $U(1)_R = 2/3$ for $H$, $\Phi_1$, $\Phi_2$,
and under which $H \Phi_2 \Phi_2$ is an allowed coupling.}
\be
W \supset y_{H33} H \Phi_3\Phi_3+y_{332}\Phi_3\Phi_3\Phi_2+y_{333}\Phi_3\Phi_3\Phi_3.
\label{eq:toyW}
\ee
In particular, note that there is no tree level $H \Phi_2 \Phi_2$ coupling in Eq. (\ref{eq:toyW}). Nonetheless this coupling is
generated at 2 loop level from the diagrams of figure 5.
\begin{figure}[ht]
\begin{center}
\includegraphics[width=14cm]{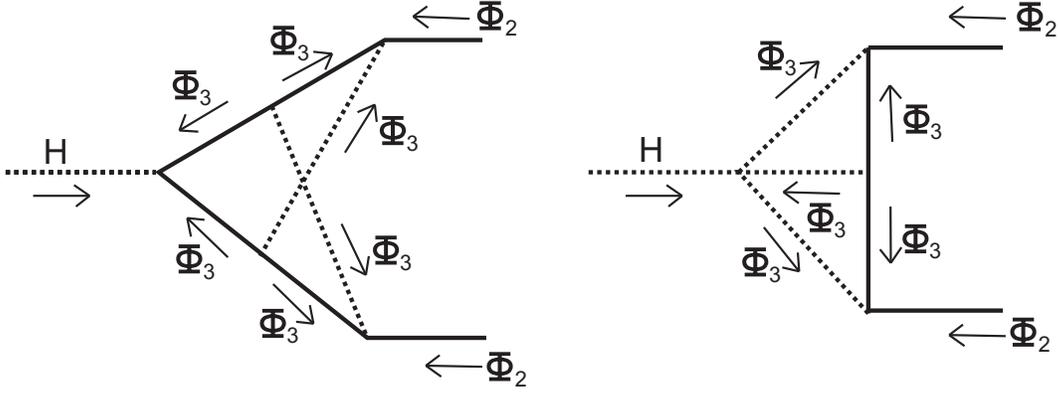}
\caption{The two graphs that generate the $H \Phi_2 \Phi_2$ Yukawa, with arrows denoting the chirality of the relevant
field. The scalar interaction in the second diagram comes from the
$\vert F_H \vert^2$ term in the Lagrangian.}
\end{center}
\label{NewYukawaGraph}
\end{figure}
This coupling is given by
\be
y_{H22} = 3 \zeta(3) \frac{y_{H33}(y_{332} y_{333})^2}{(16 \pi^2 )^2}.
\ee
The ratio of tree-level and radiative couplings is
\be
\frac{y_{H33}}{y_{H22}}= \frac{1}{3 \zeta(3)} \left(\frac{16 \pi^2}{y_{332}y_{333}}\right)^2.
\ee
The perturbativity requirement is that $\frac{y_{333}}{4 \pi}, \frac{y_{332}}{4 \pi} \ll 1$. Clearly this is 
most easily achieved for
large hierarchies, which in the Standard Model range from $\mc{O}(10^5)$  for $m_t/m_u$ to $\mc{O}(10)$ for $m_{\tau}/m_{\mu}$.

We now want to apply the above mechanism to the Supersymmetric Standard Model.
We recall the MSSM superpotential is given by
\be
W_{MSSM}=y_u \bar{u}_R Q H_u + y_d \bar{d}_R Q H_d + y_e \bar{e}_R L H_d+\mu H_u H_d,
\ee
where $y_u,y_d,y_l$ are 3x3 matrices in flavour space that describe the Yukawa coupling between the several chiral superfields $u_R,d_R,Q,e_R,L,H_u,H_d$.

Experiment tells us that the third generation is more massive than the remaining two: $m_t\gg m_c, m_u$; $m_b\gg m_s, m_d$ and $m_\tau \gg m_\mu, m_e$. The mass hierarchies within the SM are determined by the structure of the Yukawa matrices $y_u,y_d$ and $ y_l$. It is therefore reasonable to approximate them to
\be
y_u\approx\left( \begin{array}{ccc}
0 & 0 & 0 \\
0 & 0 & 0 \\
0 & 0 & y_t \end{array} \right), \,
y_d\approx\left( \begin{array}{ccc}
0 & 0 & 0 \\
0 & 0 & 0 \\
0 & 0 & y_b \end{array} \right), \,
y_l\approx\left( \begin{array}{ccc}
0 & 0 & 0 \\
0 & 0 & 0 \\
0 & 0 & y_\tau \end{array} \right).
\ee

In this section we take the Yukawa matrices to have exactly rank one and investigate whether the radiative generation of
the first two generation Yukawas can successfully describe the known mass hierarchies of the Standard Model. This will require the
introduction of new particles not present in the Standard Model.

\subsection{Quarks}
\label{subsecquarks}

Let us start by looking at the second and third generations of the up-type quarks. Given a top quark Yukawa,
the charm Yukawa can be generated radiatively via the two loop graph illustrated in figure 6.\\
\begin{figure}[h!]
\begin{center}
\includegraphics[width=14cm]{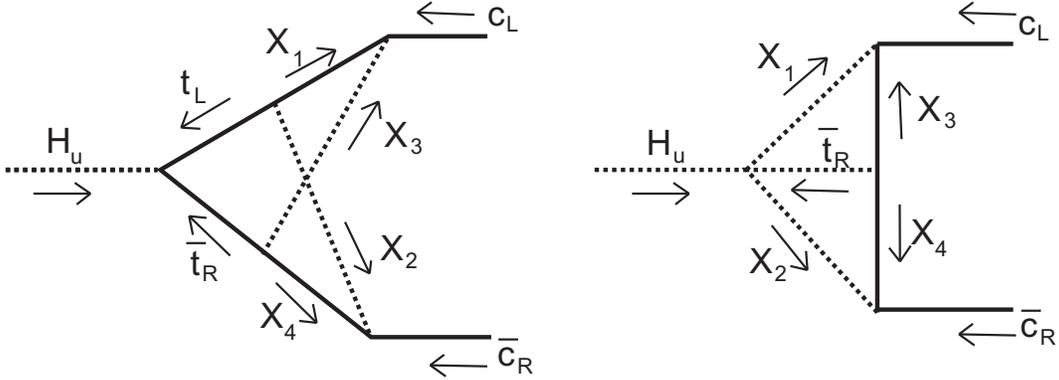}
\caption{Graphs generating a radiative charm Yukawa from the top Yukawa. There is an additional very similar graph that is not shown, in which
the 4-point scalar vertex $H_u \bar{t}_R X_1^{*} X_2^{*}$ is replaced by the vertex $H t_L X_3^{*} X_4^{*}$.}
\end{center}
\label{UpGraph}
\end{figure}

This requires the MSSM superpotential to be extended by the following terms
\be
W\supset Q^3_{L} X_{1} X_{2}+ \bar{t}_{R}X_{3} X_{4}+Q^2_{L} X_{1} X_{3}+\bar{c}_{R} X_{2} X_{4},
\label{newsuper}
\ee
where we have introduced new chiral superfields $X_{1},X_{2},X_{3} $ and $X_{4}$. These are constrained by gauge invariance. Recall that under the SM $SU(3)_c\times SU(2)_l \times U(1)_y$ the SM fields transform as shown in table  \ref{tab:SMquantumNumbers}.
\begin{table}[htbp]
  \centering
  \begin{tabular}{@{} | cc | @{}}
    \hline
    Superfield &  $SU(3)_c\times SU(2)_l \times U(1)_y$  \\
    \hline
    $Q_L^i$ & $(3,2,1/6)$ \\
    $\bar{u}_{R,i}$ & $(\bar{3},1,-2/3)$\\
    $\bar{d}_{R,i}$ &$(\bar{3},1,1/3)$\\
    $H_u$& $(1,2,1/2)$\\
    $H_d$& $(1,2,-1/2)$\\
    \hline
  \end{tabular}
  \caption{Representations of the SM gauge group, i is the family index and runs from 1 to 3.}
  \label{tab:SMquantumNumbers}
\end{table}

Gauge invariance of the superpotential (\ref{newsuper}) implies that the new fields transform under the SM gauge group as shown in table \ref{tab:IJKL}. We show three possible charge assignments for these fields. As the new fields enter only in a loop there are many
possibilities coming from the flexibility of how charge can flow around the loop.
\begin{table}[htbp]
  \centering
  \begin{tabular}{@{} | c c| c| c | @{}}
    \hline
 Superfield &  \multicolumn{3}{c}{$SU(3)_c\times SU(2)_l \times U(1)_y$} \vline\\
	\hline &A &B&C\\
    \hline
    $X_1$ & $(\bar{3},2,y)$ &$(3,2,y)$& $(1,2,y)$\\
    $X_2$ & $(1,1,-1/6-y)$&$(3,1,-y-1/6)$& $(\bar{3},1,-y-1/6)$ \\
    $X_3$ & $(1,1,-1/6-y)$&$(3,1,-y-1/6)$ &$(\bar{3},1,-y-1/6)$ \\
    $X_4$ & $ (3,1,y+5/6)$ &$ (1,1,y+5/6)$&$ (\bar{3},1,y+5/6)$ \\
    \hline
  \end{tabular}
  \caption{New fields necessary to give mass to the second generation up-type quarks}
  \label{tab:IJKL}
\end{table}
The fields $X_1, \ldots X_4$ are chiral and the spectrum of table \ref{tab:IJKL} is anomalous. The simplest way to resolve this is to extend the 
spectrum by including the conjugate partners $\bar{X}_1, \ldots \bar{X}_4$ of the
fields $X_1 \ldots X_4$ vectorlike, without adding any extra terms beyond the superpotential (\ref{newsuper}). 
This will also turn out to be necessary to avoid having extra massless states in the spectrum after supersymmetry
is broken. Also note that the $X_{2}$ and $X_{3}$ superfields have the same quantum numbers, and in principle can be identified.

In order to generate a rank 3 Yukawa matrix, there are two possibilities. The simplest is to repeat the same 2-loop mechanism used for the charm
to give mass to the up quark. However it is not sufficient simply
 to couple $u_{L}$ and $\bar{u}_{R}$ to $X_{1},X_{2}, X_3, X_{4}$ in the same way as was done for $c_{L}$ and $\bar{c}_{R}$.
 This would only generate a mass term  for the combination $u+c$ and still yield a rank 2 matrix.
 Instead we can take a further copy of the $X$ fields to generate an independent coupling of the Higgs to up-type quarks.
 The full superpotential for the up quark sector would then be
\be
\begin{split}
W\supset  \lambda t_{L} \bar{ t}_{R} H_{u}^{0}+ &  \lambda_2 \left( t_{L} X^{2}_{1} X^{2}_{2}+ \bar{t}_{R}X^{2}_{3} X^{2}_{4}+c_{L} X^{2}_{1} X^{2}_{3}+\bar{c}_{R} X^{2}_{2} X^{2}_{4} \right) +\\
& \lambda_1 ( t_{L} X^{1}_{1} X^{1}_{2}+ \bar{t}_{R}X^{1}_{3} X^{1}_{4}+u_{L} X^{1}_{1} X^{1}_{3}+\bar{u}_{R} X^{1}_{2} X^{1}_{4} ),
\end{split}
\label{eq:Wup}
\ee
where $X^{1}_{i}$ and  $X^{2}_{i}$ are the extra fields necessary to generate the 2-loop Yukawa couplings for the up and charm quarks respectively.

In principle all the trilinear terms in Eq. (\ref{eq:Wup}) can have distinct couplings, however for the sake of simplicity in our analysis we consider only two distinct new couplings: $\lambda_{1(2)}$ representing the Yukawa coupling between the first (second) generation quark superfields and the new fields. The Yukawa matrix for the up type quark sector is then:
\be \mathbf{Y}_{u} =\frac{ 3 \zeta(3)\lambda}{(4\pi)^{4}}\left( \begin{array}{ccc} |\lambda_{1}|^{4} & 0 & 0\\
0 & |\lambda_{2}|^{4}  & 0 \\ 0 & 0  & \frac{(4 \pi)^{4}}{3 \zeta(3)} \end{array} \right),
\label{eq:Yu}
\ee
where $\lambda$ is the tree level coupling of the top quark to the Higgs.
As a consequence of having only two new coupling parameters, the Yukawa matrix is given by an overall complex coupling, $\lambda$, multiplied by a real matrix.

There is a second option for generating a rank-3 Yukawa. As $X_2$ and $X_3$ have the same charges, we can consider extending
the superpotential (\ref{newsuper}) to the form
\be
W \supset \lambda H t_L \bar{t}_{R} + X_1 Q_L^i X_2^j A_{ij} + X_4 \bar{U}_R^i X_2^j B_{ij}.
\label{ewok}
\ee
Here we idenify the $X_{2}^{i}$ and $X_{3}^{i}$ fields and allow the index $j=1,2$, with $X_4^{1}=X_4^{2} \equiv X_4$, $X_1^{1}=X_1^{2} \equiv X_1$.
This implies that $A$ and $B$ are $3\times2$ matrices. This structure allows for general couplings between the quark sector and the $X$ fields. As before, at two loops this lifts the rank 1 Yukawa
structure via the diagrams of figure 6. By considering the couplings of (\ref{ewok}) it is easy to see that the
two loop Yukawa has the form
\be
Y_{ij} = \lambda \left( A_{3p}^{*} B_{jp} \right) \left( B_{3q}^{*} A_{iq} \right) \left( \frac{3 \zeta(3) }{(4 \pi)^4} \right).
\label{eq:TwoLoopY}
\ee
Both diagrams of figure 6 give contributions to the Yukawa of this index structure.
As this is of the form $v_i w_j$ it clearly only provides one new non-zero eigenvalue.

However at four loops diagrams of the form of figure 4 will occur. Examining these we see that the matrix
structure of the diagram gives a contribution to Yukawa couplings of the form
\be
Y_{ij} \sim \frac{\lambda}{(4\pi)^{8}} \left ( A^{*}_{3m} B_{qm} B_{qr}^{*} A_{ir} \right) \left( B_{3l}^{*} A_{pl} A_{ps}^{*} B_{js} \right).
\label{eq:FourLoopY}
\ee
Although this is also of the form $v^{'}_i w^{'}_j$, the structure is different from the 2-loop case and so generates a third
non-zero eigenvalue, which can also be checked numerically.
 In this case it naturally follows that the first generation will be hierarchically lighter than the second,
as the first generation Yukawa is generated at a higher loop order than that of the second generation.

The same mechanism explained above can be used to generate the mass hierarchy for the down type quarks. Looking just at the two heaviest generations we write the following superpotential
\be
\label{downsup}
W\supset H_{d}^{0} b_{L}\bar{b}_{R} + b_{L} Y_{1} Y_{2} +\bar{b}_{R} Y_{3} Y_{4}+ c_{L} Y_{1} Y_{3}+\bar{c}_{R} Y_{2} Y_{4},
\ee
to generate the mass term for the strange quark starting with a exact rank one Yukawa matrix. As before, the new chiral superfields $Y_{1},Y_{2},Y_{3},Y_{4}$ are constrained by gauge invariance to transform under the SM gauge group as given in Table \ref{tab:PQRO}.

\begin{table}[htbp]
  \centering
  \begin{tabular}{@{} | c c| c| c | @{}}
    \hline
 Superfield &  \multicolumn{3}{c}{$SU(3)_c\times SU(2)_l \times U(1)_y$} \vline\\
 \hline
	&A &B&C\\
    \hline
    $Y_1$ & $(\bar{3},2,y')$ &$(3,2,y')$& $(1,2,y')$\\
    $Y_2$ & $(1,1,-1/6-y')$&$(3,1,-y'-1/6)$& $(\bar{3},1,y')$ \\
    $Y_3$ & $(1,1,-1/6-y')$&$(3,1,-y'-1/6)$ &$(\bar{3},1,-y'-1/6)$ \\
    $Y_4$ & $ (3,1,y'-1/6)$ &$ (1,1,y'-1/6)$&$ (\bar{3},1,y'+5/6)$ \\
    \hline
  \end{tabular}
  \caption{New fields necessary to give mass to the second generation down-type quarks}
  \label{tab:PQRO}
\end{table}

 We give three possible combinations: A, B and C, all of which are defined up to the specification of the hypercharge $y'$. The pattern found here closely resembles the one needed to generate trilinear couplings in the up sector. In particular we also find that only three fields are necessary to generate a rank 2 Yukawa matrix, since $Y_2$ and $Y_3$ share the same quantum numbers and can be taken to be the same field.

As for the up-type quarks we can obtain a
rank 3 Yukawa matrix either by doubling the number of extra fields and duplicating the structure of Eq. (\ref{downsup}), or by
extending Eq. (\ref{downsup}) to allow the most general possible couplings of the $Y_2, Y_3$ fields. In this case
masses for the second and first generation quarks are generated at two loop and, at most, four loop.

\subsection{Charged and Neutral Leptons}
\label{subsecleptons}

The mass spectrum in the lepton sector is less hierarchical than the spectra of the up and down type quarks. As pointed out 
at the start of section 3, the larger the hierarchy in the mass spectrum the deeper into the perturbative regime the model will lie, which indicates that there might be some tension between the radiative generation of lepton masses and the perturbativity of the theory.  We nonetheless proceed to describe how lepton masses can originate if we start with a tree level coupling between the Higgs and the tau.

Introducing the chiral superfields $Z_{1},Z_{2},Z_{3},Z_{4}$, transforming under the SM gauge group as shown in table \ref{tab:LeptonFields}, we can extend the superpotential to include the following terms
\be
W\supset \tau_{L} Z_{1} Z_{2}+ \bar{\tau}_{R} Z_{3} Z_{4}  +  \mu_{L} Z_{1} Z_{3}+ \bar{\mu}_{ R} Z_{2} Z_{4}.
\ee

\begin{table}[htbp]
  \centering
  \begin{tabular}{@{} | c c| c| c | @{}}
    \hline
 Superfield &  \multicolumn{3}{c}{$SU(3)_c\times SU(2)_l \times U(1)_y$} \vline\\
 \hline 
	&A &B&C\\
    \hline
    $Z_1$ & $(1,2,y')$ &$(3,2,y'')$& $(\bar{3},2,y'')$\\
    $Z_2$ & $(1,1,-1/2-y'')$&$(\bar{3},1,-y''+1/2)$& $(3,1,1/2-y'')$ \\
    $Z_3$ & $(1,1,-1/2-y'')$&$(\bar{3},1,-y''+1/2)$& $(3,1,1/2-y'')$ \\
    $Z_4$ & $ (1,1,y''-3/2)$ &$ (3,1,y''-3/2)$&$ (\bar{3},1,y''-3/6)$ \\
    \hline
  \end{tabular}
  \caption{New fields necessary to give mass to the leptons}
  \label{tab:LeptonFields}
\end{table}
In this case there is no requirement the new fields should be charged under $SU(3)$.
Once we consider a further copy of these extra fields coupled to the first generation lepton fields, the resulting Yukawa matrix for the lepton sector has the same structure as that found for the up and down type quark sectors, Eq. (\ref{eq:Yu}). As before we have assumed that there are only three couplings: $\lambda$, $\lambda_{1}$ and $\lambda_{2}$.

One might also consider generic couplings between the charged leptons and the chiral superfields $Z$. By identifying $Z_2^i \equiv Z_3^i$, $Z_1^i \equiv Z_1$ and $Z_4^i \equiv Z_4$ we may write the following superpotential
\be
W \supset \lambda H^0_d \tau_L \bar{\tau}_{R} + Z_1 L^i Z_2^j A_{ij} + Z_4 \bar{E}_R^i Z_2^j B_{ij},
\label{eq:Wleptons}
\ee
where $j\in \{1,2\}$. This is analogous to Eq. (\ref{ewok}), therefore the Yukawas will be generated at two and four loop order, as in Eqs. (\ref{eq:TwoLoopY}) and  (\ref{eq:FourLoopY}). So just as in the quark sector this model will give rise to three hierarchical Yukawa couplings in the lepton sector by introducing 4 extra chiral superfields charged under the SM gauge group.

The size of quark and lepton Yukawa couplings implies a certain tension between perturbativity and radiative flavour generation.
One way to alleviate this is to relax the requirement that the two light generations have radiatively generated masses, and instead allow
only the lightest generation to be radiatively generated from the third generation. This can weaken both perturbativity and flavour constraints,
the former due to the larger hierarchy and the latter due to the weaker constraints in the $(31)$ sectors compared to the $(21)$ sector.
This mechanism can be applied to generation of rank 3 Yukawa matrices in models of branes at toric singularities \cite{Krippendorf:2010hj} which typically yield rank 2 flavour structure.

Another possible application avoiding perturbativity issues is to Dirac neutrino masses, where the required Yukawa couplings are
extremely small. As Dirac neutrino masses come from the coupling to $H_u$, this requires starting with up-type quarks
and generating all neutrino masses radiatively from the top quark coupling. Consider the superpotential
\be
W\supset A_{i j }t_L Z_1^i Z_2^j+B_{i j} \bar{t}_R Z_3^i Z_4^j+C_{i j k} n_L^i Z_1^jZ_3^k+D_{i j k} \bar{n}_R^i Z_2^jZ_4^k,
\label{eq:Wneutrinos}
\ee
Here $Z_1^{i}, Z_2^{j},Z_3^{j},Z_4^{j}$ are new fields with $i,j = 1,2 \ldots n$. We require $n \ge 2$ to generate three independent radiative
Yukawas.

The two loop Yukawa matrix describing the coupling of the neutrinos to $H^0_u$ is given by
\be
Y_{ij}= 3 \zeta(3) \frac{\lambda_t}{(4 \pi)^ 4}\sum_{m,n,p,q=1}^3 A_{mn}B_{pq}C_{npi}D_{qmj},
\ee
where $\lambda_t$ is the tree level coupling of the top quark to the Higgs field. By rotating the $Z^i$
we can diagonalise $A_{ij}$ and $B_{ij}$. For convenience we assume these take the form
$A_{ij} = A \delta_{ij}$ and $B_{ij} = B \delta_{ij}$, giving
\be
Y_{ij} = 3 \zeta(3) \frac{\lambda_t A B }{ (4 \pi)^4} \sum_{m,n=1}^3 C_{nmi} D_{mnj}.
\label{neutmassAB}
\ee

\section{The Non-Supersymmetric Phase}
\label{subsecpheno}

The previous sections have analysed the generation of radiative Yukawa couplings in the limit of unbroken supersymmetry.
However any viable real-world model must involve broken supersymmetry. How will the effect described in this paper be modified
by broken supersymmetry?

First, the models above require the addition of new fields $Z_i$. 
In the supersymmetric limit these fields are chiral and massless. Such new chiral massless fields are not viable: they lead to gauge anomalies and are inconsistent with observations. The simplest way
to address this is to make all the new fields vectorlike by including conjugate partners $\bar{Z}_i$. This clearly makes the
spectrum non-anomalous. The fields $Z_i$ cannot obtain masses via the Higgs mechanism as they do
not, by construction, couple at tree level to the Higgses. 
However, once supersymmetry is broken the fields $Z_i$ can then become massive with masses around the
weak scale via the Giudice-Masiero mechanism (assuming gravity mediation).

Secondly, in the supersymmetric limit the effect was evaluated at zero momentum, with $s=0$ in the diagrams of section 2.
The effect arises as a $\frac{0}{0}$ term, with the $0$ in the numerator from the on-shell external fields and the $0$ in the denominator
from the masslessness of the loop particles.
However, once susy is broken and the Higgs acquires a vev, it is no longer appropriate to evaluate Yukawa couplings in the deep infrared.
The relevant energy scale is instead set by the vev of the Higgs, $\langle v_H \rangle = 246 \hbox{GeV}$.
Once supersymmetry is broken, both denominator and numerator are no longer zero: the numerator is set by the Higgs vev, and the
denominator by the mass scale of the loop particles.

Another way to look at this: in the supersymmetric case, the vertex renormalisation is present in the 1PI action
 both at zero momentum (as in the calculation above) and also in off-shell processes, for example by computing 4-point functions
 (for example see \cite{West2, CGP}).
Considered off-shell at a scale $s$, the effect will remain so long as
loop particles have masses $m^2 \ll s$ and will decouple and switch off in the limit $m^2 \gg s$.

For supersymmetry to solve the hierarchy problem, its mass scale must be qualitatively close to the weak scale.
There is then a suppression by a factor of $\left( \frac{v_H}{m_{SUSY}} \right)^n$ compared to the calculation in the purely supersymmetric limit.
It is however not easy to make this calculation precise, as this requires an extension of the 2-loop calculation of 
section 2 to broken supersymmetry, massive particles and the limit $s \neq 0$. In order to make some contact between the supersymmetric models described in the previous section and the physics below the electroweak scale, where the Yukawa couplings give rise to fermionic masses, we will incorporate the effects of massive extra fields by multiplying the radiatively generated Yukawa couplings by a function $F(v_H,m_{SUSY})$. Knowledge of the full form of $F$ is equivalent to solving the loop integrations with massive fields and is beyond the scope of this paper. We shall instead discuss the structure of the Yukawas in a more qualitative fashion.

\subsection{Quark Sector}
\label{subsect:QuarkSector}

After SUSY breaking the radiatively generated Yukawa couplings will be rescaled by a factor $F(v_H/M_{SUSY})$. This implies that the simplest model in the quark sector, defined by Eq. (\ref{eq:Yu}) will become
\be
 \mathbf{Y}_{u} =\frac{ 3 \zeta(3)\lambda}{(4\pi)^{4}}\left( \begin{array}{ccc} F(v_H/M)|\lambda_{1}|^{4} & 0 & 0\\
0 & F(v_H/M) |\lambda_{2}|^{4}  & 0 \\ 0 & 0  & \frac{(4 \pi)^{4}}{3 \zeta(3)} \end{array} \right).
\label{eq:Yu2}
\ee
At the electroweak scale one can identify the eigenvalues of the Yukawa matrix with the masses of the quarks. Noting that the same model can be applied to the down quark sector we summarize the bounds on the couplings in table

\begin{table}[htdp]
\begin{center}
\begin{tabular}{c|c|c}
&Up Sector&Down Sector\\
\hline
$F\left(\frac{v_H}{M}\right)\left(\frac{|\lambda_1|}{4\pi}\right)^4$ &$4\times10^{-6}$&$3\times10^{-4}$\\
\hline
$F\left(\frac{v_H}{M}\right)\left(\frac{|\lambda_2|}{4\pi}\right)^4$  &$2\times10^{-3}$&$7\times10^{-3}$\\
\end{tabular}
\end{center}
	\caption{Scale of the couplings necessary to describe the quark mass hierarchy.}
\label{tab:couplings}
\end{table}%
We can see from table \ref{tab:couplings} that the effect\footnote{we intuitively expect $F(v_H/M)\sim\left(\frac{v_H}{M}\right)^n$ with $n>0$ and $M\sim\mc{O}(TeV)$.} of $F(v_H/M)$ is to drive the tree level couplings $\lambda_1, \lambda_2$ to higher values, increasing the tension between the radiative generation of Yukawa couplings and perturbativity of the theory in its supersymmetric phase. This also confirms that the higher the hierarchy the deeper into the perturbative regime the couplings will be as can be seen by comparing the first and second lines of table  \ref{tab:couplings}.

However a model in which both up and down quark sectors have the structure of Eq. (\ref{eq:Yu2}) cannot fully capture the flavour physics of the standard model as it yields a diagonal CKM matrix. With that in mind we turn our attention to the more generic model defined by the superpotential (\ref{ewok}). As in the simpler case analysed above, at the electroweak scale one expects both two and four loop contributions to the Yukawa matrix to receive a further suppression factor $F(v_H/M)$:\footnote{Since the four loop integral can be thought as the  product of two two loop graphs we assume that at the electroweak scale it is suppressed by $F^2(v_H/M)$.}
\bea
\label{eq:twoLoopYF}
Y_{ij} & = & F (v_H/M)\lambda \left( A_{3p}^{*} B_{jp} \right) \left( B_{3q}^{*} A_{iq} \right) \left( \frac{3 \zeta(3) }{(4 \pi)^4} \right), \\
\label{eq:fourLoopYF}
Y_{ij} & \sim & F^2 (v_H/M) \frac{\lambda}{(4\pi)^{8}} \left ( A^{*}_{3m} B_{qm} B_{qr}^{*} A_{ir} \right) \left( B_{3l}^{*} A_{pl} A_{ps}^{*} B_{js} \right).
\eea
Identifying the two loop eigenvalue with the second generation mass and the four loop with the first generation mass and assuming that the coupling matrices $A$ and $B$ are of the same order of magnitude $A \sim B $ one finds that for the up quark sector $\frac{A}{4 \pi} F^{1/4}\sim \mc{O} (0.2)$ and for the down type sector $\frac{A}{4 \pi} F^{1/4}\sim \mc{O} (0.4)$.  Once again the effect of the suppression factors is to drive the couplings to higher values, making perturbativity harder to achieve. We leave this issue to the side for the moment and proceed the analysis of the flavour structure of the model defined by Eq. (\ref{ewok}).

As noted above, besides describing the mass hierarchy, the model should also be able to reproduce the flavour structure of the SM, with the right amount of mixing between the three generations of quarks. This mixing is encoded in the CKM matrix. We briefly review its definition and then present an example from the model  we are analyzing.

One can relate the interaction eigenstates ($u_{L},u_{R},d_{L},d_{R}$) and the mass eigenstates  ($u'_{L},u'_{R},d'_{L},d'_{R}$) by unitary rotations $V_{L}^{u}$, $V_{L}^{d}$, $V_{R}^{u}$ and $V_{R}^{d}$ such that $u'_{L}=V_{L}^{u} u_{L}$, $d'_{L}=V_{L}^{d} d_{L}$, $\bar{u}'_{R}=V_{R}^{u\dagger} \bar{u}_{R}$ and $\bar{d}'_{R}=V_{R}^{d\dagger} \bar{d}_{R}$. By definition of the mass eigenstates we then have
\be
V_{L}^{u} \mathbf{Y}_{u} V_{R}^{u\dagger}=diag(m_u,m_c,m_t),\qquad
V_{L}^{d} \mathbf{Y}_{d} V_{R}^{d\dagger}=diag(m_d,m_s,m_b).
\ee
The CKM matrix is defined to be
\be
V_{CKM}\equiv V^{u\dagger}_{L} V^{d}_{L},
\ee
so we are primarily interested in computing the matrices $V^{u}_{L}$ and $ V^{d}_{L}$. This can be done by noting that they satisfy the following identities
\be
V_{L}^{u} \mathbf{Y}_{u}\mathbf{Y}_{u}^{\dagger}V_{L}^{u\dagger} =diag(m_u^{2},m_c^{2},m_t^{2}),\qquad
V_{L}^{d} \mathbf{Y}_{d}\mathbf{Y}_{d}^{\dagger}V_{L}^{d\dagger} =diag(m_d^{2},m_s^{2},m_b^{2}).
\ee
Experiments constrain the CKM matrix to be
\be
V_{CKM}=\left(
\begin{array}{ccc}
 1 & \epsilon & \epsilon^{3}\\
 \epsilon & 1 &  \epsilon^{2}\\
 \epsilon^{3} &  \epsilon^{2} & 1\\
 \end{array}
\right)
\label{eq:CKMexp}
\ee
with $\epsilon\sim0.2$ to leading order, where we have ignored the CP violating phase.

In the model we are analyzing, the  observed mass hierarchy follows from loop suppression with the second (first) generation being lighter then the third because its coupling to the Higgs is generated at two (four) loops. Therefore  the hierarchical structure is independent from the structure of the coupling matrices $A$ and $B$ of both up and down sectors, depending only on their overall scale. The mixing between generations on the other hand is encoded in the structure of these coupling matrices. In the absence of a theory that sets these couplings all that can be done is to give example where they are generated randomly that leads to (almost) realistic flavour physics in the quark sector. In order to provide such example one must specify $F(v_H/M)$. In what follows we have set $F(v_H/M)\sim (v_H/M)^2\sim1/16$ for a SUSY breaking scale of the order of the TeV. Note that there is nothing special about this choice for F, one could easily  find other examples with different choices of suppression factor provided we adjust the overall scale of the coupling matrices. This might worsen the issues with perturbativity of the theory in the supersymmetric phase but is otherwise not an issue.

Equations (\ref{eq:YuEx}) and (\ref{eq:YdEx}) provide an explicit example of  Yukawa matrices that generate a quark mass spectrum compatible with experimental constraints  within a tolerance of $5\%$:

\be
Y_u = \left(
\begin{array}{ccc}
 -0.000849865+0.000479574 i & 0.00510561-0.00421932 i & 0.0170817+0.00484836 i \\
 0.000153566-0.000262296 i & -0.000701238+0.00204042 i & -0.00557091+0.0014966 i \\
 -0.000877694+0.000861108 i & 0.00468652-0.00680455 i & 1.02224+0. i
\end{array}
\right)
\label{eq:YuEx}
\ee

\be
Y_d = \left(
\begin{array}{ccc}
 -0.00500076-0.0134638 i & -0.00517785-0.0169835 i & 0.0176335+0.00558354 i \\
 0.00509887-0.000488481 i & 0.00771031+0.00145941 i & -0.00396671+0.00282241 i \\
 -0.00476109-0.00401959 i & -0.00654716-0.00624693 i & 1.00831+0. i
\end{array}
\right)
\label{eq:YdEx}
\ee

These Yukawa matrices lead to the following CKM matrix

\be
V_{CKM}=\left(
\begin{array}{ccc}
 0.999997 & 0.00115407 & 0.00210382 \\
 0.000787639 & 0.978049 & 0.208375 \\
 0.00226662 & 0.208374 & 0.978047
\end{array}
\right).
\label{eq:CMKEx}
\ee
While Eq. (\ref{eq:CMKEx}) not fully compatible with the experimental bounds of Eq.(\ref{eq:CKMexp}), it has a similar structure and illustrates that it is possible accommodate a realistic CKM matrix in the model defined by Eq.(\ref{ewok})\footnote{When comparing  Eq. (\ref{eq:CMKEx}) with Eq.(\ref{eq:CKMexp}) one must keep in mind that the $(3,2)$ entry in Eq. (\ref{eq:CMKEx}) corresponds to the $(1,2)$ entry of Eq. (\ref{eq:CKMexp}), the (3,3) to the (1,1) and so on. This is due to the inner workings of the SingularValueDecomposition function of Mathematica.}.

\subsection{Lepton Sector}

The radiative generation of Yukawa couplings in the charged lepton sector, given by Eq. (\ref{eq:Wleptons}) is analogous to the quark sector model, Eq. (\ref{ewok}). Therefore the effects of supersymmetry breaking in the eigenvalues of the charged lepton Yukawa matrix will be similar, 
in particular two and four loop eigenvalues will be suppressed by the function $F(v_H/M)$ as in Eqs. (\ref{eq:twoLoopYF}) and (\ref{eq:fourLoopYF}).
Given that $m_e=511 \hbox{keV}$, $m_\mu=105 \hbox{MeV}$ and $m_\tau=1776 \hbox{MeV}$ we find that in order to describe the mass ratios one needs $\frac{A}{4 \pi} F^{1/4}\sim \mc{O}(0.4)$, assuming $A\sim B$. Once again perturbativity is an issue and it is made worse by the inclusion of $F(v_H/M)$.

We finally turn our attention to the neutrinos. Unlike the case of the quarks and charged leptons, the neutrino spectrum is not completely well defined. Data from solar and atmospheric oscillations tell us that the spectrum can be either hierarchical or quasi degenerate with $\sum_{i=1}^3 m_i \leq 0.62 \hbox{eV}.$ In what follows we show that the model defined by Eq. (\ref{eq:Wneutrinos}) can yield a neutrino mass spectrum consistent with the current experimental constraints in a region in which the couplings are perturbative.
At the electroweak scale, the two loop Yukawa couplings of Eq.(\ref{neutmassAB}) become
\be
Y_{ij} = 3 \zeta(3) F(v_H/M)\frac{\lambda_t A B }{ (4 \pi)^4} \sum_{m,n=1}^3 C_{nmi} D_{mnj},
\label{neutmassAB2}
\ee
where $\lambda_t$ is the tree level coupling of the top quark to the Higgs. Noting that $m_t=172 \hbox{GeV}$ and taking the worst case scenario for the neutrino spectrum in terms of perturbativity, $m_\nu\sim 10^ {-1} \hbox{eV}$, we find that the scale of the new couplings  $A,B,C,D$ necessary to give a Dirac mass to the neutrinos is
\be
\frac{A}{4\pi}\sim \frac{B}{4\pi}\sim \frac{C}{4\pi}\sim \frac{D}{4\pi} \sim \mc{O}(10^{-3})F^{-1/4}(v_H/M),
\label{eq:neutConst2}
\ee
which is in perturbative regime for a wide range of values for $F(v_H/M)$ . Note that this conclusion will hold independently of whether the neutrino mass spectrum
is quasi-degenerate or hierarchical.

What type of neutrino mass spectrum should we expect? From Eq.(\ref{neutmassAB2}) we see that it is not possible to make any definitive
statements as the form of the neutrino spectrum comes from the form of the couplings $C_{ijk}$ and $D_{ijk}$. Without an underlying
theory of the structure of $C$ and $D$ we cannot say what form of neutrino spectrum is expected. We make however two comments.

First, the model defined by Eq. (\ref{eq:Wneutrinos}) is capable of producing
realistic mass spectra for the neutrinos. We do so by providing two representative examples
for quasi-degenerate and hierarchical spectra. As before one must attribute a value to the function $F(v_H/M)$. We choose $F(v_H/M)=1/16$ and note that this particular choice does not affect the generality of our conclusions, impacting only in the overall scale of the new couplings as seen in Eq.(\ref{eq:neutConst2}).

\paragraph{Quasi degenerate spectrum}
Considering the couplings in the Appendix, the Yukawa matrix is
\be
\frac{Y_{ij}}{\lambda}=10^{-13}\left(
\begin{array}{ccc}
 -1.78816+0.204652 i & 6.46274-0.959414 i & 5.03087-5.57817 i \\
 -4.08909-4.97862 i & 2.70535-0.866756 i & -7.60958-2.26435 i \\
 -2.11153+4.95205 i & -0.943148+7.50286 i & 3.18599+2.0113 i
\end{array}
\right)
\ee

Upon diagonalisation, the masses of the neutrinos are
\be
m_1=0.095\: \hbox{eV}, \qquad m_2=0.139\: \hbox{eV}, \qquad m_3=0.255\: \hbox{eV}.
\ee

\paragraph{Hierarchical Spectrum}

\be
\frac{Y_{ij}}{\lambda}=10^{-13}\left(
\begin{array}{ccc}
 2.32916+6.81065 i & -0.376941-1.1337 i & 3.05031+3.21891 i \\
 -3.29302+5.79996 i & 5.10187-4.47371 i & -3.84616+3.28579 i \\
 5.11796+5.52427 i & -6.13005+5.43047 i & 9.41025+1.71153 i
\end{array}
\right)
\ee

\be
m_1=0.000579497\: \hbox{eV}, \qquad m_2=0.137975\: \hbox{eV}, \qquad m_3=0.317119\: \hbox{eV}.
\ee

Second, if we assume that $C_{ijk}$ and $D_{ijk}$ are generated randomly, then a degenerate spectrum is preferred over a hierarchical
one. Generating the elements of $C$ and $D$ as complex numbers whose real and imaginary parts have mean zero and unit variance,
we obtain numerically a probability distribution for $m_3/m_1$ as $(m_3/m_1)^{-1.8}$, disfavouring highly hierarchical spectra.
In a similar fashion, under the same assumptions for the elements of $C$ and $D$, the mixing angles are anarchic and in general large.
However, as there is no fundamental theory for the structure of the couplings $C_{ijk}$ and $D_{ijk}$, limited significance should be attached
to this.

\section{Flavour Constraints and Phenomenology}
\label{secflavour}

A necessary element of the models described above was the introduction of 
new degrees of freedom charged under the SM gauge group. These new fields will introduce a non-trivial flavour structure which can be constrained by experimental data. In this section we investigate if the new physics described in the previous sections is compatible with the known experimental limits, in particular we constrain the quark sector using flavour changing neutral currents (FCNC) and the lepton sector through the decays $\mu\rightarrow e \gamma$, $\tau\rightarrow e \gamma$ and $\tau\rightarrow \mu \gamma$.

\subsection{Quark Sector}

To quantify the effect of having new fields and couplings in the theory one assumes that these extra fields are heavier than the SM fields. This allows us to integrate them out and consider an effective field theory of the form:
\be
\mc{L}_{eff}=\mc{L}_{SM}+\Delta \mc{L},\qquad \Delta \mc{L}=\sum \frac{c}{\Lambda^{d-4}}\mc{O}^{d},
\ee
where $\mc{L}_{SM}$ denotes the SM Lagrangian and $\Delta \mc{L}$ represents all the possible dimension d operators that can be obtained by integrating out fields above a given cut off scale $\Lambda$. Of particular interest to us is the set of $\Delta F=2$ operators in $\mathcal{L}_{eff}$ of the form of $\frac{c_{sd}}{\Lambda^{2}}(\bar{s}_{R}d_{L})(\bar{s}_{L} d_{R})$.The other relevant terms are listed in table \ref{tab:FCNCbounds} together with the bounds on $\Lambda$ (when $c_{ij} =1$)  and $c_{ij}$ (when $\Lambda$ is 1TeV) \cite{Isidori:2010kg}.

\begin{table}[htbp]
  \centering
  \begin{tabular}{ | c | c c | c c | }
    \hline
Operator &  \multicolumn{2}{c|}{Bounds on $\Lambda$ in TeV} & \multicolumn{2}{|c}{ Bounds on $c_{ij}$} \vline \\
	&Re &Im&Re&Im  \\
    \hline
	$(\bar{s}_{R}d_{L})(\bar{s}_{L} d_{R})$ &$1.8\times 10^{4}$&$3.2\times 10^{5}$&$6.9\times10^{-9}$ &$2.6\times10^{-11}$\\
	$(\bar{b}_{R}d_{L})(\bar{b}_{L} d_{R})$ &$1.9\times 10^{3}$&$3.6\times 10^{3}$&$5.6\times10^{-7}$ &$1.7\times10^{-7}$\\
	$(\bar{b}_{R}s_{L})(\bar{b}_{L} s_{R})$ &\multicolumn{2}{c|}{$3.7\times10^{2}$} & \multicolumn{2}{|c}{$1.3\times 10^{-5}$} \vline \\
	$(\bar{c}_{R}u_{L})(\bar{c}_{L} u_{R})$ &$6.2\times 10^{3}$&$1.5\times 10^{4}$&$5.7\times10^{-8}$ &$1.1\times10^{-8}$\\
    \hline
  \end{tabular}
  \caption{Bounds on dimension 6 $\Delta F=2$ operators.}
  \label{tab:FCNCbounds}
\end{table}

These bounds are obtained by assuming that when a given operator is generated at tree level in the SM, the contribution arising from new physics is negligible. This allows the decoupling of new physics effects from the experimental determination of the CKM matrix. The limits are then set by requiring that the contribution from the terms in $\Delta \mc{L}$  to a particular process to be smaller than the corresponding SM contribution. For a more detailed explanation see \cite{Isidori:2010kg}.

In the model we are analysing, each of the terms in table \ref{tab:FCNCbounds} can be  mapped into the diagram of Fig. 7.
\begin{figure}[ht]
 \label{fig7}
\begin{center}
\includegraphics[width=9cm]{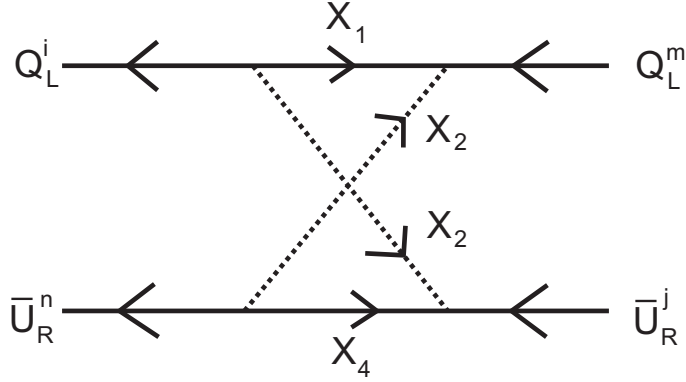}
\caption{A 4-fermion diagram that is induced by the same fields that give rise to radiative Yukawa generation.}
\end{center}
\end{figure}
We estimate the amplitude for this process to be given by
\be
\mc{A}(Q_L^i\bar{U}_R^{j*} Q^{m*}_L \bar{U}_R^n)\sim \frac{1}{(4\pi)^2 M^2} \sum_{r,s=1}^2 A_{i r}A^*_{ms}B_{ns}B^*_{jr}.
\ee
Note that in matrix notation this amplitude is proportional to $N_{ij} N^\dagger_{ms}$, where $N$ is a  $3\times 3$ matrix defined by $N\equiv A B^\dagger$. We are interested in the cases in which $n=i$ and $m=j$. There is then a direct correspondence between the amplitude $\mc{A}(Q_L^i\bar{U}_R^{j*} Q^{j*}_l \bar{U}_R^i)$ and the coefficients $c_{ij}/\Lambda^2$ constrained by table \ref{tab:FCNCbounds}:
\be
\frac{N_{ij}N^\dagger_{ij}}{(4\pi)^2 M^ 2}\equiv \frac{c_{ij}}{\Lambda^2}.
\label{eq:map}
\ee
This provides a mapping between the experimental constraints from FCNC and the trilinear couplings between the quarks and the new fields. In order for the model to succeed, the couplings necessary to explain the quark mass hierarchy must be compatible with the constrains in table \ref{tab:FCNCbounds}.
Since we are interested in having TeV scale SUSY in order to solve the hierarchy problem, we set $M\sim 1 \hbox{TeV}$ 
which simplifies Eq. ({\ref{eq:map}) to $c_{ij}=\frac{N_{ij}N^\dagger_{ij}}{(4\pi)^2}$. One can then read off the constrainst on $N$ from the second collumn of table \ref{tab:FCNCbounds}.
Note that the matrix structure of the two and four loop Yukawa couplings of  Eqs. (\ref{eq:twoLoopYF})  and (\ref{eq:fourLoopYF}) can be expressed in terms of $N$ as
\be
Y_{ij}|_{2loop}\sim N_{i3} N^*_{3j},  \qquad Y_{ij}|_{4loop}\sim (NN^\dagger)_{i3}(N^\dagger N)_{j3}.
\label{eq:YN}
\ee

Let us start by focusing on the down quark sector. The constraints of table \ref{tab:FCNCbounds} imply that
\be
N_{12}N^*_{21}<\mc{O}(10^{-3}), \qquad
N_{23}N^*_{32}<\mc{O}(10^{-7}),\qquad
N_{13}N^*_{31}<\mc{O}(10^{-5}).
\ee
which considering $N_{ij}\sim N_{ji}$ translates into
\be
N\sim \mc{O}\left(
\begin{array}{ccc}
n_{11}&10^{-3/2}&10^{-5/2}\\
10^{-3/2}&n_{22}& 10^{-7/2}\\
10^{-5/2}&10^{-7/2}&n_{33}\\
\end{array}
\right).
\label{eq:boundsN}
\ee
One then finds that the two and four loop eigenvalues, Eq. (\ref{eq:YN}), that follow from Eq (\ref{eq:boundsN}) are several orders of magnitude below the values what is necessary to describe the down quark mass hierarchy. Equivalently, it was discussed in Section \ref{subsect:QuarkSector} that the scale of the couplings needed is  $\frac{A}{4 \pi} F^{1/4}\sim\frac{B}{4 \pi} F^{1/4}\sim \mc{O} (0.4)$ which translates into $N$ several orders of magnitude larger than the bounds of Eq. ({\ref{eq:boundsN}}). It seems therefore hard to reconcile the radiative flavour generation for the down quarks with FCNC constraints.

Let us now look at the up quark sector. Due to the large mass of the top quark, the only constraint on the up quark sector involves the first and second generation. From Eq. (\ref{eq:map}) and table \ref{tab:FCNCbounds} it follows that for the up sector the sole constraint is
\be
N_{12}N^*_{21}\sim \mc{O}(10^{-6}).
\ee
It is possible to set $N_{12}\sim N_{21}\sim\mc{O}(10^{-3})$ and still find the right mass hierarchy by setting the couplings such that
\bea
N_{12}=\sum_{m=1}^2 A_{1m} B^*_{2m} < \mc{O}(10^{-3}),\\
N_{21}=\sum_{m=1}^2 A_{2m} B^*_{1m} < \mc{O}(10^{-3}).
\eea
One can go even further and set $N_{12}$ and $N_{21}$ to zero. This then translates into having trilinear couplings between the down quarks and the new fields that obey
\bea
B_{21}^*=-\frac{A_{12}}{A_{11}}B^*_{22},\\
B_{11}^*=-\frac{A_{22}}{A_{21}}B^*_{12}.
\label{eq:condAB}
\eea
Note that Eq. (\ref{eq:condAB}) does not correspond to a suppression of the couplings, instead it just requires a certain "alignment" between the coefficients of the coupling matrices $A$ and $B$. We have numerically confirmed that radiatively generated up quark spectra obeying Eq. (\ref{eq:condAB}}) can be found.

\subsection{Charged Lepton Sector}

Flavour constraints to the leptonic sector come mostly from the lepton flavour violating decays $\mu\rightarrow e \gamma$, $\tau\rightarrow \mu \gamma$ and $\tau\rightarrow e \gamma$:
\bea
\label{eq:BR21}
BR(\mu\rightarrow e \gamma)< 1.2\times10^{-11},\\
\label{eq:BR32}
BR(\tau\rightarrow \mu \gamma)<4.4\times10^{-8},\\
\label{eq:BR31}
BR(\tau\rightarrow e \gamma)<3.3\times10^{-8}.
\eea

 \begin{figure}[ht]
 \label{fig7}
\begin{center}
\includegraphics[width=11cm]{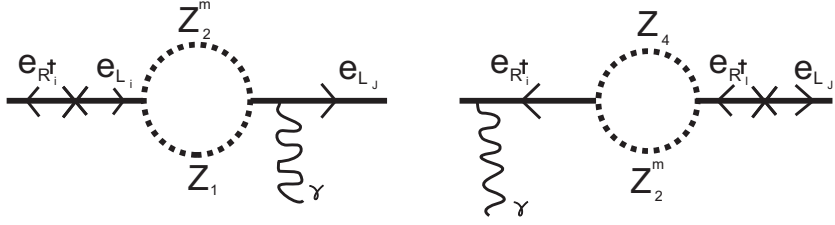}
\caption{Diagrams that can generate the dimension 5 coupling $e^\dagger_{Ri}\sigma^{\mu\nu} e_{Lj}F_{\mu\nu}$.}
\end{center}
\end{figure}

These very constrained decay modes can be generated in our model at the one loop level as depicted in Fig. 8. These diagrams correspond to dimension five operators in the effective field theory of the form $e^\dagger_{Ri}\sigma^{\mu\nu} e_{Lj}F_{\mu\nu}$ and $e^\dagger_{Li}\sigma^{\mu\nu} e_{Rj}F_{\mu\nu}$ respectively. The Wilson coefficients corresponding to these operators must have dimensions of inverse mass. We estimate the amplitude for the process I to be of the order of
\be
\mc{A}(l_i\rightarrow l_j \gamma)\sim \frac{(AA^\dagger)_{ij}}{(4\pi)^2}\frac{e m_{l_i}}{M^2},
\ee
where $e$ denotes the electric charge of the field that emits the photon. Note that  the factor of $m_{l_i}$ comes from the chirality flip in the  the incoming lepton line. The decay  width is proportional to the square of the amplitude and one also must include a factor of $1/8\pi$ from the two body phase space and 3 powers of $m_{l_i}$ to give the right dimensions. The decay width is then given by:
\be
\Gamma(l_i\rightarrow l_j \gamma)=\frac{1}{8 \pi} |\mc{A}|^2 m_{li}^3,
\ee
from which one gets the following branching ratio
\be
BR(l_i\rightarrow l_j \gamma)\sim T_{l_i} |(AA^\dagger)_{ij}|^2\frac{e^2 m_{l_i} ^5}{2 M^4 (4\pi)^5},
\label{eq:BR}
\ee
where $T_{l_i}$ is the mean life of the $l_i$ lepton, $T_{l_i}=\Gamma_{tot}^{-1}$.
It then follows that, taking the experimental bounds into account, the couplings are constrained to be
\be
|(AA^\dagger)_{ij}|^2 \lesssim \frac{ 2 (4\pi)^5 M^4}{T_{l_i}m_{l_i} ^5 e^2}BR(l_i\rightarrow l_j \gamma),
\ee
where
\be
T_\mu=2.19\times10^{-6} \hbox{s} \qquad \text{and} \qquad T_\tau=290.6\times10^{-15} \hbox{s},
\ee
and we take the experimental upper limits on the branching ratios as given by Eqs. (\ref{eq:BR21})-(\ref{eq:BR31}). One then finds that for graph I constrains the coupling matrix $A$ :
\bea
(A A^\dagger)_{12}\lesssim \mc{O}(10^{-3}),\\
(A A^\dagger)_{32}\lesssim\mc{O}(10^{-2}),\\
(A A^\dagger)_{31}\lesssim\mc{O}(10^{-2}),
\eea
up to factors of $e^2$. Recalling that to describe the mass hierarchy in the charged lepton sector the characteristic size of the couplings is $\frac{A}{4\pi} F^{1/4}\sim \mc{O}(0.4)$ one concludes that the couplings are several orders of magnitude above the constraints from lepton flavour violating decays. We note that graph II would give rise to similar constraints on the coupling matrix $B$.

All the flavour constraints above have assumed that both light generations obtain their masses radiatively from the third generation.
If the only radiative generation is $3 \to 1$, then flavour constraints are significantly weakened (for example $t \to u$ FCNCs
 are much less constrained than $c \to u$ FCNCs). 

\subsection{Phenomenology}

We close with a few short comments on phenomenology. The structure of the superpotential (\ref{eq:Wup}) implies that the new fields
$Z_i$ would couple pairwise to Standard Model degrees of freedom: there is an effective $Z_2$ symmetry under which Standard Model fields
have charge $+1$ and the new fields have charge $-1$. This implies that any collider production of the $Z$ fields would necessarily involve
a missing energy signature, with the lightest such field being stable, and also a potential dark matter candidate. 

\section{Conclusions}

In this paper we made use of the subtle distinction between one-particle irreducible and Wilsonian actions to 
 investigate field theories that generate the flavour structure of the Standard Model in the supersymmetric phase.
The well known non-renormalization theorem of the superpotential in supersymmetric theories applies only to the Wilsonian superpotential.
When interacting massless particles are present in the theory, the physical (1PI) superpotential can get renormalised, even in the absence
of any SUSY breaking effects.

We have developed models aiming at explaining the flavour structure in the supersymmetric Standard Model in which the 
observed mass hierarchies arise through radiative effects. 
This required the introduction of extra massless vectorlike chiral superfields in the supersymmetric phase.
New couplings introduced between the SM superfields and the new degrees of freedom in the supersymmetric phase allow for the breaking of the rank one Yukawa structure, generating masses for the second and first generations of quarks and charged leptons at two and four loop respectively.
 
The magnitude of this effect is modifed once supersymmetry is broken and we move below the electroweak scale. 
 To incorporate this effect into our computation we considered the loop generated couplings would be rescaled by a function $F(v_H/M)$ whose effect is to increase the strength of the new couplings between the SM fields and the extra superfields, increasing the tension between getting the observed mass hierarchy and the perturbativity of the theory.
 
An important set of constraints comes from flavour physics. For the quark sector these constraints arise from measurements of flavour changing neutral currents. These are very severe in the down quark sector, where it is observed that the allowed couplings are several orders of magnitude bellow the values necessary to explain the mass hierarchy. In the up quark sector however, since the constraints are weaker, it is possible to generate the mass hierarchy by aligning the coupling matrices in a particular way. The more stringent constraints for the charged leptons arise from upper bounds on lepton flavour violating decays. Similarly to what happens in the down quark sector, the constraints impose an upper limit on the couplings which is much smaller than that necessary to explain the leptonic mass hierarchy.

Although these precision constraints appear very severe, there may be ways to alleviate them.
For example, if the only radiative Yukawa generation was $3 \to 1$ then both perturbativity and flavour constraints are much weakened due to the
larger hierarchy and the weaker bounds on $3 \to 1$ FCNCs.
Furthermore, it may be possible to develop models based for which extra symmetries or alignment structure may weaken or eliminate the
 flavour constraints. For this reason we do not regard the ideas presented in this paper as necessarily ruled out, even if the particular models described would need further development.

For the case of Dirac neutrinos, supersymmetric radiative generation appears to provide a successful mechanism to generate masses compatible with experimental bounds. Starting from the top-Higgs Yukawa it is possible to generate both quasi-degenerate and hierarchical neutrino masses in the regime where the perturbative description holds. As a consequence of the lightness of neutrino masses, the tension with perturbativity encountered for the other quarks and leptons does not hold.

\subsection*{Acknowledgments}

JC is supported by a Royal Society University Research Fellowship and by Balliol College, Oxford.
FGP is suported by Funda\c{c}\~{a}o para a Ci\^{e}ncia e a Tecnologia (Portugal) through the grant SFRH/BD/35756/2007.
We thank Ian Jack, Tim Jones, Eran Palti and Fernando Quevedo for conversations. JC thanks the University of Glasgow for hospitality, where a preliminary
version of this work was presented.

\bibliographystyle{JHEP}

\begin{thebibliography}{10}

\bibitem{0601204}
  V.~Braun, Y.~-H.~He, B.~A.~Ovrut,
  JHEP {\bf 0604}, 019 (2006).
  [hep-th/0601204].

\bibitem{09042186}
  L.~B.~Anderson, J.~Gray, D.~Grayson, Y.~-H.~He, A.~Lukas,
  Commun.\ Math.\ Phys.\  {\bf 297}, 95-127 (2010).
  [arXiv:0904.2186 [hep-th]].

\bibitem{0302105}
  D.~Cremades, L.~E.~Ibanez, F.~Marchesano,
  JHEP {\bf 0307}, 038 (2003).
  [hep-th/0302105].

\bibitem{0404229}
  D.~Cremades, L.~E.~Ibanez, F.~Marchesano,
  JHEP {\bf 0405}, 079 (2004).
  [hep-th/0404229].

\bibitem{aiqu}
  G.~Aldazabal, L.~E.~Ibanez, F.~Quevedo, A.~M.~Uranga,
  JHEP {\bf 0008}, 002 (2000).
  [hep-th/0005067].

    \bibitem{Krippendorf:2010hj}
  S.~Krippendorf, M.~J.~Dolan, A.~Maharana and F.~Quevedo,
  JHEP {\bf 1006} (2010) 092
  [arXiv:1002.1790 [hep-th]].
  

\bibitem{Cecotti:2009zf}
  S.~Cecotti, M.~C.~N.~Cheng, J.~J.~Heckman and C.~Vafa,
  arXiv:0910.0477 [hep-th].

\bibitem{09102413}
  J.~P.~Conlon, E.~Palti,
  JHEP {\bf 1001}, 029 (2010).
  [arXiv:0910.2413 [hep-th]].

\bibitem{hepth0612110}
  S.~A.~Abel and M.~D.~Goodsell,
  JHEP {\bf 0710}, 034 (2007)
  [arXiv:hep-th/0612110].

\bibitem{09105496}
  F.~Marchesano and L.~Martucci,
  Phys.\ Rev.\ Lett.\  {\bf 104}, 231601 (2010)
  [arXiv:0910.5496 [hep-th]].

\bibitem{11021973}
  C.~P.~Burgess, S.~Krippendorf, A.~Maharana and F.~Quevedo,
  arXiv:1102.1973 [hep-th].

\bibitem{Ibanez1982}
  L.~E.~Ibanez,
  Phys.\ Lett.\  {\bf B117}, 403 (1982).

\bibitem{9601262}
  N.~Arkani-Hamed, H.~-C.~Cheng, L.~J.~Hall,
  Phys.\ Rev.\  {\bf D54}, 2242-2260 (1996).
  [hep-ph/9601262].

\bibitem{08041996}
  S.~Nandi, Z.~Tavartkiladze,
  Phys.\ Lett.\  {\bf B672}, 240-245 (2009).
  [arXiv:0804.1996 [hep-ph]].

\bibitem{09064657}
  P.~W.~Graham, S.~Rajendran,
  Phys.\ Rev.\  {\bf D81}, 033002 (2010).
  [arXiv:0906.4657 [hep-ph]].


\bibitem{West1}
  P.~C.~West,
  Phys.\ Lett.\  B {\bf 261} (1991) 396.

\bibitem{West2}
    P.~C.~West,
  Phys.\ Lett.\  B {\bf 258} (1991) 375.

\bibitem{JackJonesWest}
  I.~Jack, D.~R.~T.~Jones and P.~C.~West,
  Phys.\ Lett.\  B {\bf 258} (1991) 382.
  
\bibitem{DunbarJackJones}
  D.~C.~Dunbar, I.~Jack, D.~R.~T.~Jones,
  Phys.\ Lett.\  {\bf B261}, 62-64 (1991).

\bibitem{CGP}
  J.~P.~Conlon, M.~Goodsell, E.~Palti,
  JHEP {\bf 1011}, 087 (2010).
  [arXiv:1007.5145 [hep-th]].



\bibitem{Grisaru:1979wc}
  M.~T.~Grisaru, W.~Siegel, M.~Rocek,
  Nucl.\ Phys.\  {\bf B159 } (1979)  429.

\bibitem{Seiberg:1993vc}
  N.~Seiberg,
  Phys.\ Lett.\  B {\bf 318} (1993) 469
  [arXiv:hep-ph/9309335].

\bibitem{Seiberg:1994bp}
  N.~Seiberg,
  arXiv:hep-th/9408013.


\bibitem{Shifman:1986}
  M.~A.~Shifman and A.~I.~Vainshtein,
  Nucl.\ Phys.\  B {\bf 277} (1986) 456
  [Sov.\ Phys.\ JETP {\bf 64} (1986) 428]
  [Zh.\ Eksp.\ Teor.\ Fiz.\  {\bf 91} (1986) 723].

\bibitem{KL}  
  V.~Kaplunovsky, J.~Louis,
  Nucl.\ Phys.\  {\bf B422}, 57-124 (1994).
  [hep-th/9402005].

\bibitem{ArkaniHamedMurayama}
  N.~Arkani-Hamed, H.~Murayama,
  JHEP {\bf 0006}, 030 (2000).
  [hep-th/9707133].


\bibitem{Tkachov:1981wb}
  F.~V.~Tkachov,
  Phys.\ Lett.\  B {\bf 100} (1981) 65.

\bibitem{Jones:1982zf}
  D.~R.~T.~Jones and J.~P.~Leveille,
  Nucl.\ Phys.\  B {\bf 206}, 473 (1982)
  [Erratum-ibid.\  B {\bf 222}, 517 (1983)].

\bibitem{Buchbinder1}
  I.~L.~Buchbinder, S.~M.~Kuzenko and A.~Y.~Petrov,
  Phys.\ Lett.\  B {\bf 321} (1994) 372.
  
  \bibitem{Buchbinder2}
  I.~L.~Buchbinder, S.~Kuzenko, Z.~.Yarevskaya,
  Nucl.\ Phys.\  {\bf B411}, 665-692 (1994).

\bibitem{PeskinSchroeder}
  M.~E.~Peskin, D.~V.~Schroeder,
  Reading, USA: Addison-Wesley (1995) 842 p.
  

\bibitem{Isidori:2010kg}
  G.~Isidori, Y.~Nir and G.~Perez,
  arXiv:1002.0900 [hep-ph].




\end{thebibliography}

\end{document}